\documentclass{article}

\usepackage{times}
\usepackage{epsfig}
\usepackage{graphicx}
\usepackage{amsmath}
\usepackage{amssymb}
\usepackage{caption}
\usepackage{subcaption}
\usepackage{algorithm}
\usepackage{algorithmic}
\usepackage{hyperref}
\usepackage{geometry}
\begin{document}

\title{Learning to Recommend via Meta Parameter Partition}

\author{Liang Zhao, Yang Wang, Daxiang Dong, Hao Tian\\
Baidu Research Institute\\
Sunnyvale, CA 94089, USA\\
\texttt{\{zhaoliang07, wangyang59\}@baidu.com}
}

\date{\vspace{-5ex}}

\maketitle

\begin{abstract}
In this paper we propose to solve an important problem in recommendation 
-— user cold start, based on meta leaning method. Previous meta learning
approaches finetune \textit{all} parameters for each new user, which is both computing and storage expensive. In contrast, we divide model parameters into
fixed and adaptive parts and develop a two-stage meta learning algorithm to learn them separately.
The fixed part, capturing user invariant features, is shared by all users and is learned during offline meta learning stage. The adaptive part, capturing user specific features, is learned during online meta learning stage.   
By decoupling user invariant parameters from user dependent parameters,
 the proposed approach is more efficient and storage cheaper than previous methods. 
It also has potential to deal with catastrophic 
forgetting while continually adapting for streaming coming users.
 Experiments on production data demonstrates that the 
proposed method converges faster and to a better performance
 than baseline methods. Meta-training without online 
meta model finetuning increases the AUC from 72.24\% 
to 74.72\% (2.48\% absolute improvement). Online meta training
 achieves a further gain of 2.46\% absolute improvement 
comparing with offline meta training. 
\end{abstract}

\section{Introduction} 
Personalized recommender systems are playing more and 
more important roles in web-based and mobile applications. 
The goal of learning to recommend
 is to learn a training paradigm that takes as input a set of 
items from a user's history and generates a function or user 
model that can be applied to new items and to predict how 
likely this user will click that item -- click through rate(CTR) prediction. 
One of the key challenges faced by conventional Matrix 
Factorization~\cite{linden2003:mf,koren2008:mf} 
methods is to make personalized recommendation for new 
users arriving sequentially -- user cold start problem. 

A common way to solve cold start problem is by leveraging information 
from other users to help with model training. For example, 
McMahan et al.~\cite{mcmahan2017:fl} propose Federated 
learning by using a unified model for all users which is joint 
trained with all user data. The limitation with such approach 
is that the learned model is biased toward major interests and 
may not reflect personal interests of new users due to lacking
 training data for them. 
Meta learning algorithms such as MAML~\cite{finn2017:maml} provide a promising
 way to learn a good initialization for new task/user models.
However, previous meta learning approaches~\cite{chen2018:recsys} finetune 
\textit{all} parameters for each new user, which is both computing and storage expensive. 

In this paper, we propose to divide model parameters into
fixed and adaptive parts, and
we develop a two-stage meta learning algorithm to learn them separately.
The fixed part, capturing user invariant features, is shared by all users and is learned during offline meta learning stage. The adaptive part, capturing user specific features, is learned during online meta learning stage.
By decoupling user invariant parameters from user dependent parameters,
 the proposed approach is more efficient and storage cheaper than previous methods.
It also has potential to deal with catastrophic
forgetting while continually adapting for streaming coming users.

We evaluated the proposed algorithm on the real-world 
problem of news feed recommendation. The production 
dataset was collected from our news recommender platform. 
We trained the model with around 595k records from 57k 
users and evaluated on 17k records from 2k new users. Each 
training and test users have only 10 and 8 records per session 
in average, respectively. This is a typical few-shot learning setting.
 The experimental results demonstrate 
that the proposed method converges faster and to a better 
performance than baseline methods. Meta-training without 
online meta model finetuning increases the AUC from 
72.24\% to 74.72\% (2.48\% absolute improvement). Our algorithm 
continuous to improve with additional training examples/
iterations during online meta training and achieves 
a further gain of 2.46\% absolute improvement over offline 
meta training. 

Our main technical contributions include three aspects: 
First, we divide model parameters into fixed and adaptive parts for
capturing user invariant and user dependent features separately. 
Second, we developed an offline and an online meta learning algorithms to 
learn fixed and adaptive parameters, respectively. 
Third, we evaluated our algorithm on production data 
and demonstrates its advantages over conventional methods. 

\section{Related Work} 
Our work is closely related to CTR prediction, few-shot 
learning, meta learning, online learning and continual learning. 
In this section, we provide a brief summary of related 
work in these areas. 

\textbf{CTR prediction:} In recently years, several deep learning-
based approaches~\cite{liu2015:recsys,zhang2014:recsys, 
zhang2016:recsys,qu2016:recsys,cheng2016:recsys,guo2017:recsys} have been
 applied to CTR prediction in recommdendation systems.
 Deep neural network (DNN) has potential to 
learn expressive features for accurate CTR prediction. CNN-
based models ~\cite{liu2015:recsys} are biased to the interaction 
between neighboring features. RNN-based models~\cite{zhang2014:recsys} are suitable for data with sequential dependency. 
FNN-based model~\cite{zhang2016:recsys} is 
limited by the capacity of the pretrained factorization machine 
(FM). PNN~\cite{qu2016:recsys} introduces product-
layer between embedding and fully-connected layers. Wide-
Deep~\cite{cheng2016:recsys} combines wide and deep models 
where wide-part relies on feature engineering. DeepFM~\cite{guo2017:recsys} models both low and high order feature 
interaction in an end-to-end manner. More recently reinforcement 
learning based methods~\cite{zhao2018page,zhao2018list, 
zheng2018rl,chen2019rl} have been proposed for 
modeling user behavior and item recommendation. All these 
methods do not handle cold-start problem very well. In contrast, 
our approach can handle users with fewer records very 
well by using online meta learning. 

\begin{figure}[t]
\centering
\vspace{-0.3 em}
\centerline{\includegraphics[width= 1.0 \textwidth]{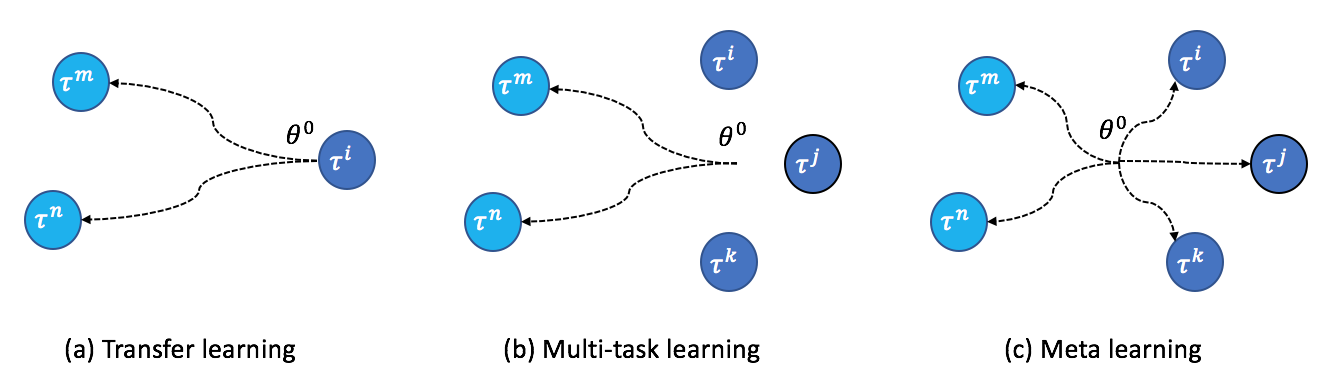}}
\vspace{-0.3 em}
\caption{Compare different learning methods for model parameter initialization (dark blue ones are source tasks and light blue
ones are target tasks): (a) transfer learning (b) multi-task learning (c) meta learning. This illustrates that meta learning provides
a better initialization for finetuning with fewer iterations than both transfer learning and multi-task learning.}
\vspace{-0.5 em}
\label{fig:meta-init}
\end{figure}

\textbf{Few-shot learning:} The tremendous gains in DNN relies 
on large amount of labeled training data. However, in 
many situations such as cold-start problem in recommendation 
system, the labeled training data for new users are very 
limited. This requires to solve the problem with few-shot 
learning~\cite{lake2015few} techniques. 
One popular way to deal with limited training data is 
transfer learning~\cite{pratt1993few} which pretrains the model on a 
source task with large amounts of data and finetunes on a target 
task with small amounts of data. An extension of transfer 
learning is multi-task learning~\cite{caruana1997mtl} when a set of 
source tasks is available for pretraining the model. In contrast 
to these few-shot learning methods, we employ meta 
learning for transferring knowledge shared among users. As 
shown in Figure~\ref{fig:meta-init}, meta learning generates a better model parameter 
initialization than both transfer learning and multi-
task learning. 

\textbf{Meta learning for recommendation:}
Meta learning \cite{schmidhuber1987:meta, finn2017:maml, nichol2018:meta} provides an effective way to perform transfer learning across users by means of shared parameters \cite{vartak2017:recsys}, shared model \cite{chen2018:recsys}, or by shared initialization \cite{chen2018:recsys}. All these methods assume that the training data are available at once and train the model in batch mode. In contrast, our approach works in a more realistic scenario where training data arrives in a sequence.
 Recently, \cite{finn2019online} proposes an online meta learning algorithm for processing sequential data and show benefits on several computer vision tasks.
A drawback with these meta learning
approaches is that they finetune \textit{all} parameters for each new user, which is both computing and storage expensive. In contrast, our approach enables all user models share user-invariant parameters which saves both storage and time for adapting.

\textbf{Continual learning:}
Our problem setting is related to continual learning \cite{thrun1998lifelong, zhao1996incremental}. Continual learning with neural network has been explored in recent years.
Based on the way for overcoming catastrophic forgetting, current methods can be classified into approaches with fixed model size and with increasing model size. Approaches with fixed model size either employ certain constraints \cite{kirkpatrick2017ewc, aljundi2019ssl} to control parameter changes when learning new concepts/tasks or use additional memory \cite{lopez2017gem, rebuffi2017icarl, aljundi2018mas, aljundi2019online} to store certain information about previous data/tasks for retraining purpose. Approaches with increasing model size \cite{li2019learn} learn new tasks with added small amount of parameters while keeping the previously learned parameters fixed. 
 Unlike previous work, our approach decouples 
user invariant parameters from user dependent parameters and has potential to deal with catastrophic forgetting by keeping user invariant parameters fixed and only updating user dependent parameters. 

\textbf{Online learning:}
Similar to the setting of continual learning, online learning handles a sequential setting with streaming tasks. One early work is Follow the Leader(FTL) \cite{hannan1957online} followed with various improvements \cite{cesa2006online, hazan2006online, shalev2012online}. We cast meta learning in an online learning setting and derived an efficient online meta learning algorithm that captured the practice of online learning and leads to promising experimental results.

\section{Problem Formulation} 
A common way of building personalized recommender systems
 is to estimate the click through rate (CTR) of a user
-- the probability that the user will click on a recommended
item, and use the CTR to rank the items for displaying to
the user. Therefore, accurate CTR prediction is important
for recommender systems.

\subsection{Personalized CTR Prediction Problem}
We formulate CTR prediction as a binary classification problem. 
The training data is a set of $n$ instances $(X, y)$, where 
$X$ is an m-field (or m-slot) data record usually consisting of 
a pair of user and item attributes, and $y$ indicates whether 
the user click the item or not. $y = 1$ means that the user click the 
item, and $y = 0$ otherwise. The problem is to learn a function 
or user model parameterized with $\theta$ for predicting the 
probability of a user clicking a specific item: 

$$
\hat{y} = f_{CTR}(x; \theta), 
$$
where $x = [x_{slot_1}, ..., x_{slot_m}]$. 

Conventional recommendation system uses a single unified 
user model for predicting a new user's CTR and the 
model is trained with all user data. In contrast, personalized 
recommendation system has a model for each user, which is 
trained with each user's own data. When the training data is 
few for a new or occasional user, the learned model will be 
overfitting to the training data. Meta learning provides a way 
to handle few-shot learning as shown below. 

\subsection{Meta Learning for Model Initialization} 

We consider learning a user model for predicting the user 
CTR as a task and cast personalized CTR prediction in a 
meta learning setting. The underlying idea of meta learning 
is to use a set of source tasks $\{\tau^{1}, ... \tau^{k}\}$ 
to learn a meta 
model whose parameters is used to initialize the parameters 
of each user model (see Figure~\ref{fig:meta-init}(c)). 
This meta initialization 
enables learning new task requires fewer training data and 
steps than transfer learning and multi-task learning as shown 
in Figure~\ref{fig:meta-init}. 

In the context of personalized CTR prediction, we use a 
set of users with large amounts of records to train a meta 
model and then initialize each new user model with the meta 
model parameters for online finetuning. Let the loss of a 
sampled task be $L_{\tau}$. 
 The objective function is to minimize 
the expected loss after updating the task model parameter 
for k steps using function $U_{\tau}^k(\theta)$: 

\begin{equation}
\theta^{*} = \arg \min_{\theta} E_{\tau}[L_{\tau}(U_{\tau}^k(\theta))]
\end{equation}

\begin{figure}[t]
\centering
\vspace{-0.3 em}
\centerline{\includegraphics[width= 0.6 \textwidth]{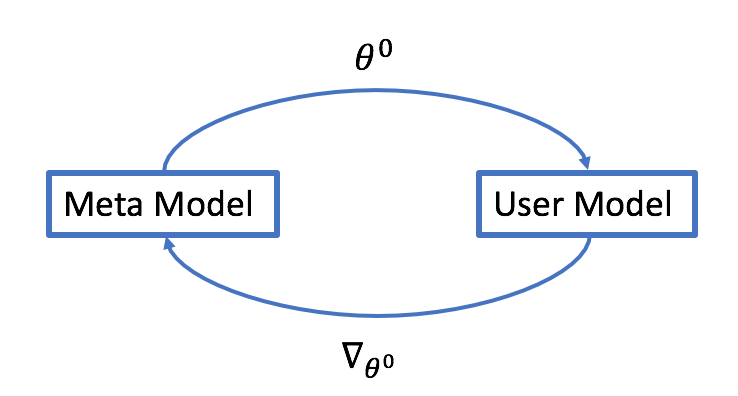}}
\vspace{-0.3 em}
\caption{Meta learning for personalized user modeling. The
meta model is used to initialize the user model with $\theta^0$ 
and the user model provides meta-gradient $\nabla_{\theta^0}$ 
for updating the meta model.}
\vspace{-0.5 em}
\label{fig:meta}
\end{figure}

We employ Reptile~\cite{nichol2018:meta} 
-- a simplified MAML~\cite{finn2017:maml} algorithm -- for calculating the meta gradient $\nabla_{\theta^0}$. 

\begin{equation}
\nabla_{\theta^0} = \frac{(\tilde\theta - \theta^0)} {\epsilon}
\end{equation}

where
\begin{equation}
\tilde\theta = U_{\tau}^{k}(\theta)
\end{equation}

and $\epsilon$ is a hyperparameter controlling the stepsize. Function 
$U_{\tau}^{k}(\theta)$ 
is an operator that updates the user model parameter $\theta$ for $k$ 
steps of SGD or Adam~\cite{kingma2015adam} using data sampled from $\tau$. 

This formulation enables online meta-learning without 
distinctive training and test stages, and makes it a more natural 
choice in our online continual adaptation setting. 

\section{Two-Stage Meta Learning Algorithm} 

Our approach contains a meta model and a user model for 
each user. The meta model is used to initialize the user 
model with $\theta^0$  
and the user model provides meta-gradient $\nabla_{\theta^0}$ 
for updating the meta model (see Figure~\ref{fig:meta}). Our training 
consists of two stages: offline meta training on the source 
tasks and online meta training and test on the target tasks. 
During offline learning stage, the meta model is trained to 
capture user invariant features and to provide a good initialization
 for each user model. During online learning stage, 
each user model is trained for adapting personal features to 
changing data distribution while keeping user invariant features
 captured by meta model fixed. By decoupling user invariant
 parameters from user dependent parameters, the proposed
 approach not only saves storage and time for adapting, but also can 
deal with catastrophic forgetting while 
continually adapting to user specific parameters. The following
 subsections describe the two stages in detail.

\subsection{Offline Meta Learning An Initialization} 
The goal of offline meta learning is to train the meta model 
for capturing user invariant features and to provide a good 
initialization for each user model. Let $\theta^0$ 
be the parameters of meta model and $\theta^{t}$ ($t \in [1, n])$ 
be the parameters of task/user $\tau_t$. Algorithm 1 summarizes the offline
 meta learning procedure which is adopted from Reptile~\cite{nichol2018:meta} 
 -- a simplified MAML. 

\begin{algorithm}
   \caption{Offline Meta Learning Algorithm}
   \label{alg:offline}
\begin{algorithmic}
   \STATE {\bfseries Input:} Innerloop steps $N$, meta loop steps $M$
   \STATE Initialize $\theta^{0}$
   \FOR{$step=1$ {\bfseries to} $M$}
   \STATE Sample $n$ tasks from source tasks
   \FOR{each new task $\tau_t$}
   \STATE Initialize $\theta^{t}$ with $\theta^{0}$
   \STATE $\theta^{t} \leftarrow SGD(\theta^{t}; D_{train}^t, \alpha)$
   \ENDFOR
   \STATE $\theta^{0} \leftarrow \theta^{0} - \beta (\theta^{0} - \theta^{t})$
   \ENDFOR
\end{algorithmic}
\end{algorithm}

The offline meta learning procedure consists of two loops. 
The outer loop updates the meta model parameters and the 
inner loop updates the training task parameters. In the beginning
 of the algorithm, the meta model is initialized with random
 noise. In each inner loop, a task is sampled from source 
task set and the task model parameters are initialized with 
meta model parameters. Then $k$ batches of data are sampled 
from the task for updating the task model parameters with $k$
SGD steps. In each outer loop, the meta model is updated 
based on meta-gradient calculated with Eq.(2). 

\subsection{Online Meta Learning for Personalized Finetuning} 
In online setting, both users and user click records arrive
 sequentially. We derive an online meta learning algorithm
 to update each user's model incrementally as training 
data available once the user either click or skip the recommended
 item. Here we use the meta model learned in the 
offline training stage to initialize each user model and fine-
tune the embedding and classifier layers while keeping the
rest layers fixed. Therefore, we split the model parameters
into two groups: a fixed one and an adaptive one as follows
$\theta = [\theta_{fixed}, \theta_{adaptive}]$. 
Algorithm 2 summarizes the online meta learning algorithm. 

\begin{algorithm}[htb]
   \caption{Online Meta Learning Algorithm}
   \label{alg:online}
\begin{algorithmic}
   \STATE {\bfseries Input:} Innerloop steps $N$, meta loop steps $M$
   \STATE Initialize $\theta^{0} = [\theta^{0}_{fixed}, \theta^{0}_{adaptive}]$ with meta parameters learned in the offline stage
   \FOR{$step=1$ {\bfseries to} $M$}
   \FOR{each new task $\tau_t$ arriving sequentially from target tasks}
   \STATE Initialize $\theta^{t}$ with $\theta^{0}$
   \FOR{each new item arriving sequentially in the task}
   \STATE obtain the click or not label and use it as a training data
   \STATE $\theta_{adaptive}^{t} \leftarrow SGD(\theta_{adaptive}^{t}; D_{train}^t, \alpha)$
   \ENDFOR
   \ENDFOR
   \STATE $\theta_{adaptive}^{0} \leftarrow \theta_{adaptive}^{0} - \beta (\theta_{adaptive}^{0} - \theta_{adaptive}^{t})$
   \ENDFOR
\end{algorithmic}
\end{algorithm}

The online meta learning procedure consists of two loops. 
The outer loop updates the adaptive meta model parameters
 while keeping the user invariant layers fixed; the inner 
loop updates the task/user specific parameters. In the beginning 
of the algorithm, the meta model is initialized with parameters
 learned in the offline training stage. In each inner loop, 
for each task arriving sequentially from the target task set, 
the task model parameters are initialized with meta model 
parameters. Then for each new item arriving sequentially 
in the task, the click-or-not label is obtained and the 
task specific model parameters are updated with $k$ 
SGD steps. In each outer loop, the adaptive parameters $\theta_{adaptive}^0$ 
of meta model is updated based on meta-gradient calculated with Eq.(2). 

\section{Experiments} 
\subsection{Production Dataset} 
\begin{figure}[t]
\begin{center}
\begin{subfigure}[b]{0.4 \textwidth}
\includegraphics[width=\textwidth]{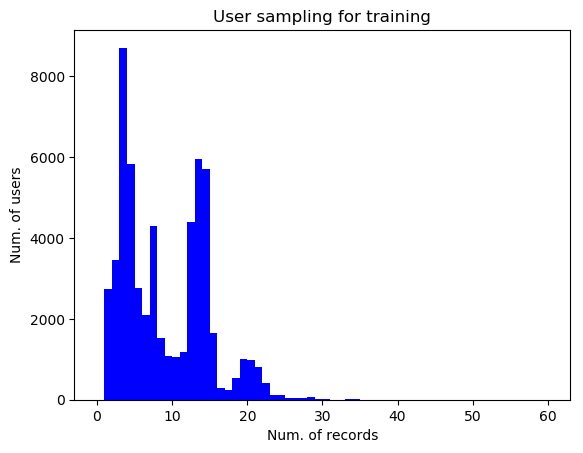}
\caption{\label{sfig:a}}
\end{subfigure}
\begin{subfigure}[b]{0.4 \textwidth}
\includegraphics[width=\textwidth]{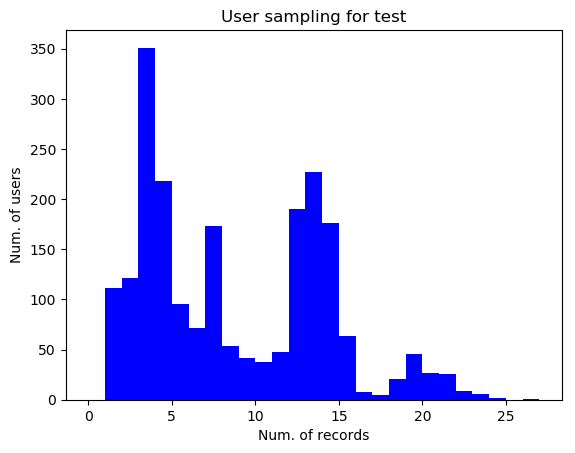}
\caption{\label{sfig:b}}
\end{subfigure}
\end{center}
\caption{ Statistics of News Feed dataset
(a) Histogram distribution of number of records per user in training 
data 
(b) Histogram distribution of number of records per user in test 
data 
}
\label{fig:data}
\end{figure}

We collected the production datasets from news feed distributed
 systems for two hours. The data from the first hour 
is used for offline meta training and those from the second 
hour for online meta training and testing. Figure~\ref{fig:data} shows the 
histogram distributions of the number of records per user in 
training and test stages. Both training and test data have four 
modes (see Figure~\ref{fig:data}) corresponding to four types of users: 
active users with around 20 records per session, regular users 
with around 13 records per session, occasional users with 
around 7 records per session, and new users with around 3 
records per session. 

\begin{table}[thb]
\centering
\caption{Summary of News Feed dataset}

\label{tab:data}
\begin{tabular}{|l|c|c|}
\hline
Data & Train & Test \\
\hline
Records & 595140 & 17731 \\
\hline
Users & 57184 & 2020 \\
\hline
Average records/user & 10 & 8 \\
\hline
\end{tabular}
\end{table}

Table~\ref{tab:data} summarizes the collected data, from which we 
can see that although there are around 595k training data, 
for each user there are only 10 records in average for training
-- which is a typical few-shot learning setting. There is 
no overlap between training and test users. Therefore, this 
dataset can test both the generalization and fast adaptation 
capabilities of the proposed algorithm.

\subsection{DNN Architecture} 
\begin{figure}[t]
\centering
\vspace{-0.3 em}
\centerline{\includegraphics[width= 0.8 \textwidth]{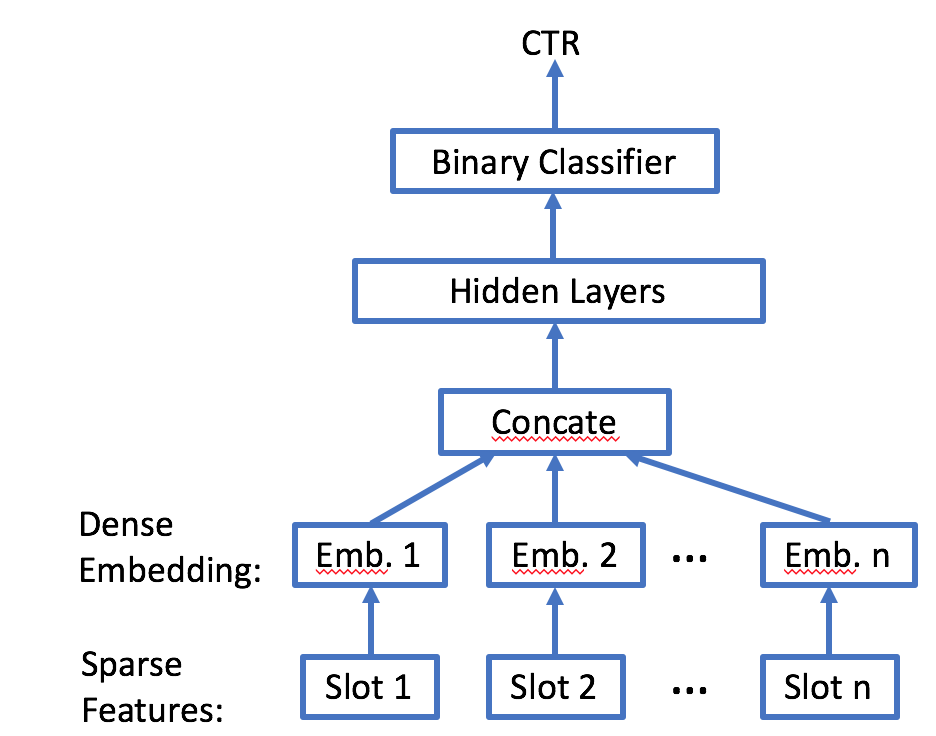}}
\vspace{-0.3 em}
\caption{DNN architecture for user modeling}
\label{fig:dnn}
\vspace{-0.5 em}
\end{figure}

For fair comparison, all algorithms use the same neural network 
architecture (see Figure~\ref{fig:dnn}) for training. The network consists 
of one embedding layer for each input slot feature 
and three fully connected hidden layers followed by a softmax layer 
for binary classification. The input consists of 571 
slots, encoding the user features(such as age, 
gender, location) and news item features(such as item-id, 
title-term, item-category, author-id). Each slot attribute 
is encoded with an embedding of 16 dimension. The three 
hidden layers contain 128, 64, 32 hidden units, respectively. 
The output layer contains two units for binary classification. 

\subsection{Parameter Settings} 
\begin{figure}[h]
\begin{center}
\begin{subfigure}[b]{0.4 \textwidth}
\includegraphics[width=\textwidth]{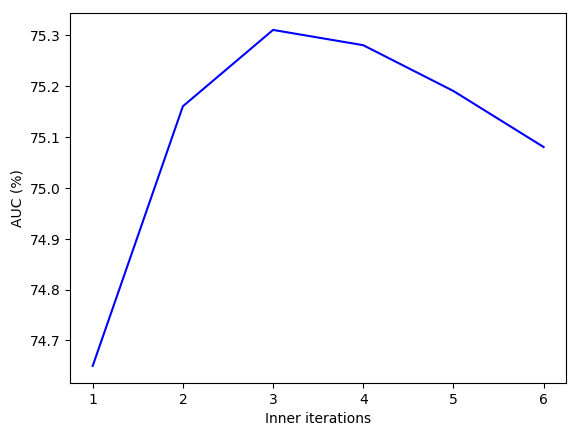}
\caption{\label{sfig:a}}
\end{subfigure}
\begin{subfigure}[b]{0.4 \textwidth}
\includegraphics[width=\textwidth]{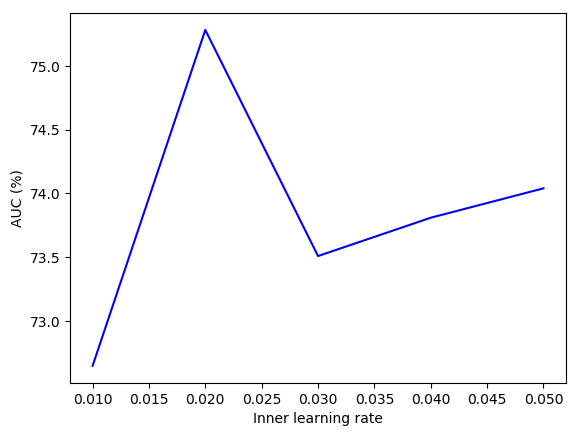}
\caption{\label{sfig:b}}
\end{subfigure}
\end{center}
\caption{Results for hyperparameter tuning 
(a) Results for tuning the number of inner loop iterations 
(b) Results for tuning the inner loop learning rate}
\label{fig:tune}
\end{figure}

We trained the neural networks with around 595k records 
collected from 57k users and evaluated on 17k records from 2k new 
users. For our experiments, we used vanilla SGD in both 
outer (meta model updating) and inner (user model updating) 
loops. The hyperparameters in our algorithm is fine-
tuned based on those reported in Reptile. Figure~\ref{fig:tune} provides the results of hyperparameter 
tuning which indicates that when setting the inner 
iteration to be 3 and learning rate to be 0.02, the proposed 
algorithm achieves the highest AUC. 

\begin{table}[thb]
\centering
\caption{Hyper-parameters for meta-learning algorithm}
\label{tab:para}
\begin{tabular}{|l|c|}
\hline
Parameter & Value \\
\hline
Inner learning rate & 0.02 \\
\hline
Inner batch size & 4 \\
\hline
Inner iterations & 5 \\
\hline
Outer learning rate & linearly annealed from 1.0 to 0.0 \\
\hline
Outer batch size during training & 5 \\
\hline
Outer iterations & 100k \\
\hline
\end{tabular}
\end{table}

Table~\ref{tab:para} lists the hyperparameters for our meta learning 
algorithm. 

\subsection{Experimental Results} 

\begin{table}[thb]
\centering
\caption{Experimental results on Feeds dataset}

\label{tab:sum}
\begin{tabular}{|l|c|}
\hline
Method  & AUC (mean $\pm$ std) \\ \hline 
base & 72.24 $\pm$ 0.21 \\ \hline
base+finetune & 73.53 $\pm$ 0.11 \\ \hline 
meta & 74.72 $\pm$ 0.09 \\ \hline 
proposed & \textbf{77.18 $\pm$ 0.09} \\ \hline
\end{tabular}
\end{table}
 
We compared the proposed algorithm with three baselines 
by running each method five times. The three baseline methods are: \textit{base}, \textit{base+finetune}, and \textit{meta}. \textit{base} is the 
method with a unified user model trained with SGD algorithm. 
For fair comparison, its training data includes the training 
part in the meta testing phase. \textit{base+finetune} is a hybrid 
algorithm with the same pretraining phase as the base algorithm and 
an online training/test phase as a meta algorithm. \textit{meta} is 
the conventional meta learning algorithm.

We use area under ROC curve(AUC) as evaluation metric in
our experiments. Table~\ref{tab:sum} summarizes 
the experimental results (mean and standard deviation 
of five runs) on our production dataset. From Table~\ref{tab:sum} we can 
see that the proposed method performs significantly better 
than three baselines with an increase of AUC from 72.24\% 
to 77.18\% (4.94\% absolute improvement). Among the three 
baseline methods, \textit{base} algorithm with a unified model performs 
significantly worse than the other methods. Conventional 
meta learning method performs better than two other 
baselines with an increase of AUC from 72.24\% to 74.72\% 
(2.48\% absolute improvement). Note that a small improvement 
in AUC evaluation for example 0.275\% is likely to lead 
to a significant increase on CTR prediction accuracy(for example 
3.9\% improvement in this case). The performance of \textit{base+finetune} 
falls between \textit{base} and \textit{meta}. In particular, both \textit{meta} and 
\textit{base+finetune} perform better than \textit{base}, showing that the 
user's own data is more effective than the data from other 
users for training. Both \textit{meta} and the proposed algorithms 
perform better than two \textit{base} methods demonstrating that 
the initialization model trained with meta approach is better than conventional joint training a unified model due to 
meta learning's better generalization ability. The proposed 
two-stage meta learning algorithm further improves the performance 
of conventional meta learning algorithm by 2.46\% 
absolute increase of AUC due to online meta learning. 

\subsection{Ablation Study} 
We conducted ablation study to find out the contribution 
from different factors: meta initialization, decoupled fine-
tuning, and online meta learning. 

\begin{figure}[t]
\centering
\vspace{-0.3 em}
\centerline{\includegraphics[width= 0.8 \textwidth]{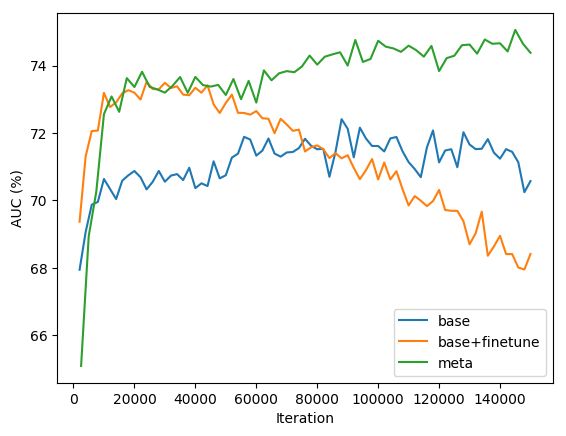}}
\vspace{-0.3 em}
\caption{Comparing meta learning algorithm with two baselines}
\label{fig:exp}
\vspace{-0.5 em}
\end{figure}

The first experiment we conducted is to compare the conventional 
meta learning method with two baselines. 
From Figure 6 we can see that \textit{base} algorithm did not adapt 
to new user data, and \textit{base+finetune} tends to overfitting to 
training data. In comparison, \textit{meta} converged faster and to 
a higher AUC, which demonstrates that meta learning does 
provide a better initialization than two baselines.

The second experiment we conducted is to evaluate which 
network layers should be fixed during online meta learning. 

\begin{table}[thb]
\centering
\caption{Experimental results on fixing different network layers}

\label{tab:fix}
\begin{tabular}{|l|c|}
\hline

Fixed layers & AUC \\ \hline 
1st (emb) & 74.49 \\ \hline
2nd (hid) & 74.72 \\ \hline
3rd (hid) & 74.72 \\ \hline
4th (hid) & 74.71 \\ \hline
5th (clf) & 74.70 \\ \hline
2nd, 3rd & \textbf{74.73} \\ \hline
2nd, 3rd, 4th & 74.71 \\ \hline
\end{tabular}
\end{table}

Table~\ref{tab:fix} reports the experimental results on fixing different 
network layers, from which we can see that fixing the embedding 
layer (1st layer) resulted in worst performance and fixing the 
middle two hidden layers (2nd and 3rd layers) achieved the 
highest performance. This means that both embedding layer 
and classifier layer should be finetuned in order to adapt 
to changing of user attributes and user interests, while the 
hidden layers are user invariant parameters and can be kept 
fixed. 

\begin{figure}[t]
\centering
\vspace{-0.3 em}
\centerline{\includegraphics[width= 0.8 \textwidth]{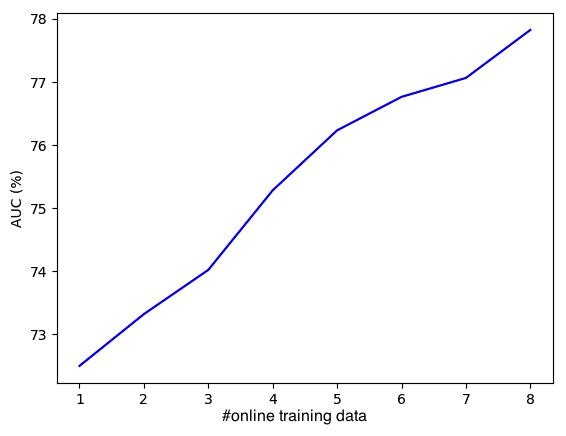}}
\vspace{-0.3 em}
\caption{Results for online meta learning with increasing
number of training data}
\label{fig:num}
\vspace{-0.5 em}
\end{figure}

The third experiment we conducted is to evaluate the gain 
we can get from online meta learning by continually updating 
both meta and user model with training data online. 
Figure~\ref{fig:num} demonstrates that our algorithm improves consistently 
with additional training examples/iterations during online 
meta learning stage. 

\section{Conclusion} 
In this paper, we examined an import problem in recommendation -- user cold start and introduced a two-stage meta leaning algorithm to solve it. The differences with other meta learning methods are:
(1) we combine an offline meta and an online meta model to get better initialization for streaming coming users;
(2) in online meta model, we divide model parameters into fixed and adaptive parts. In this way, the model can capture cross-user invariant and adapt to personal features.
Previous meta learning
approaches finetune all parameters for each new user, which is both computing and storage expensive. In contrast, our approach enables all user models share user-invariant parameters which saves both storage and time for adapting. 
Experiments on production data demonstrates 
that the proposed method converges faster and to a better 
performance than baseline methods. Meta-training without 
online finetuning increases the AUC from 
72.24\% to 74.72\% (2.48\% absolute improvement). Our algorithm 
keeps on improving with additional training examples/iterations 
during online meta training and achieves a further gain of 2.46\% 
absolute improvement comparing with offline meta training.
In our future work, we will further study how the proposed approach can 
deal with catastrophic forgetting while continually adapting to user specific
parameters, benefiting from decoupling user invariant parameters from user dependent parameters.


\end{document}